\documentclass[pra,aps,twocolumn,superscriptaddress,showpacs,eqsecnum]{revtex4}

\usepackage{amsmath}
\usepackage{graphicx}
\usepackage{color}
\usepackage{hyperref}
\usepackage{ulem}

\begin{document}

\title{Coherent control of light transport in a dense and disordered atomic ensemble}

\author{A.S. Sheremet}
\affiliation{Department of Theoretical Physics, St-Petersburg State
Polytechnic University, 195251, St.-Petersburg, Russia}
\affiliation{Russian Quantum Center, Novaya 100, 143025 Skolkovo, Moscow Region, Russia}
\author{D.F. Kornovan}
\affiliation{St-Petersburg Academic University RAS, 194021, St.-Petersburg, Russia}
\author{L.V. Gerasimov}
\affiliation{Department of Theoretical Physics, St-Petersburg State
Polytechnic University, 195251, St.-Petersburg, Russia}
\author{B. Gouraud}
\address{Laboratoire Kastler Brossel, UPMC-Sorbonne Universit\'es, CNRS, ENS-PSL Research University, Coll\`ege de France, 4 place
Jussieu, 75005 Paris, France}
\author{J. Laurat}
\address{Laboratoire Kastler Brossel, UPMC-Sorbonne Universit\'es, CNRS, ENS-PSL Research University, Coll\`ege de France, 4 place
Jussieu, 75005 Paris, France}
\author{D.V. Kupriyanov}
\email{kupr@dk11578.spb.edu}
\affiliation{Department of Theoretical Physics, St-Petersburg State
Polytechnic University, 195251, St.-Petersburg, Russia}
\affiliation{Russian Quantum Center, Novaya 100, 143025 Skolkovo, Moscow Region, Russia}

\date{\today}

\begin{abstract}
Light transport in a dense and disordered cold atomic ensemble, where the cooperation of atomic dipoles essentially modifies their coupling with the radiation modes, offers an alternative approach to light-matter interfacing protocols. Here, we show how the cooperativity and quasi-static dipole interaction affect the process of light propagation under the conditions of electromagnetically-induced transparency (EIT). We perform comparative analysis of the self-consistent approach with ab-initio microscopic calculations and emphasize the role of the interatomic interaction in the dipoles' dynamics. Our results show that in such a dense and strongly disordered system the EIT-based light storage protocol stays relatively insensitive to configuration variations and can be obtained with essentially less atoms than it is normally needed for dilute systems.
\end{abstract}


\maketitle

\section{Introduction}
\noindent Coherent control of light propagation through a cold and optically deep atomic ensemble has been the basis for a variety of remarkable light-matter interfacing protocols, including single-photon or entanglement storage in Raman- or EIT-based quantum memories with applications to quantum information networks \cite{ReviewMemory1,FL,Sanders,ReviewMemory2,CDLK, OAM, Segment, Kimble}. In these experiments, and in most associated theoretical studies, relatively low densities and dilute configurations are considered \cite{FIM,PSH}. Interestingly, with higher density - i.e. up to one atom in a volume of radiation wavelength - the effective interface between light and matter and reliable light storage can be obtained with essentially less number of atoms than it is achieved in dilute gases. In the diffraction limit, for a given optical depth $b_0$, the minimal number of required atoms scales indeed as $N\gtrsim b_0^2/(n_0\lambdabar^3)$, where $n_0$ is the density of atoms and $\lambdabar=k^{-1}$ is the inverse wave number of the radiation field \cite{Sheremet}. This estimate can even lead to a smaller number of participating atoms if the light is transversally confined into a nanophotonic waveguide, e.g. in tapered nanofibers where the divergence associated with Fraunhofer diffraction vanishes \cite{nanofiber1,nanofiber2}. Such one-dimensional configuration is highly appealing due to its wide range of potential applications \cite{Chang,Mitsch}.

However for atomic systems in the limit of high density and strong disorder, when $n_0\lambdabar^3\sim 1$, the cooperative dynamics and static interatomic interactions essentially modify  the scattering process and can lead, for instance, to localization phenomena in the light transport \cite{Akkermans,SKKH,Bellando,Sheremet}. These collective effects can also be important when the transport is controlled by external coherent fields. In this paper, we therefore investigate the basic process of coherent Raman control of a signal pulse entering such a medium. We present comparative analysis of the problem for $\Lambda$-type atoms. Our detailed study is based on the self-consistent macroscopic Maxwell description and on the alternative ab-initio microscopic calculation of the scattering process. Our results, confirmed by both calculation schemes and based on atomic arrays with small number of atoms, indicate that the transport stays relatively insensitive to configuration variations.

\section{Self-consistent approach}
\begin{figure}[tp]
{$\scalebox{0.5}{\includegraphics*{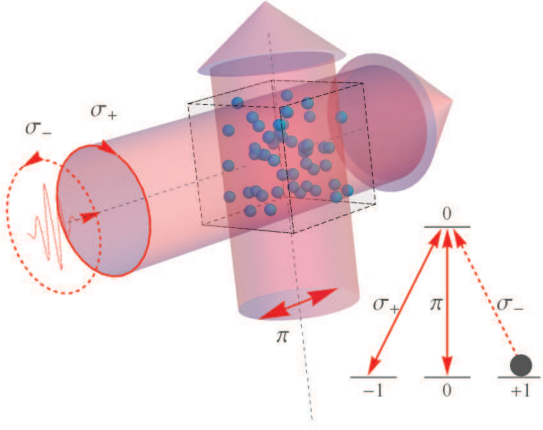}}$ }
\caption{(Color online) Excitation scheme of a dense and disordered atomic ensemble in the presence of control modes. The atoms have a total spin angular momentum $F_0=1$ in the ground state and $F=0$ in the excited state. A weak probe beam with $\sigma_-$ polarization propagates along the sample. The two control modes with equal Rabi frequencies $\Omega_c$ address the empty adjacent transitions with orthogonal polarizations $\sigma_+$ and $\pi$.}
\label{fig1}%
\end{figure}

The energy level diagram and excitation scheme are shown in Fig.~\ref{fig1}. We consider a dense ensemble of $\Lambda$-type atoms with the minimal accessible number of quantum states, i.e. with angular momentum $F_0=1$ in the ground state and $F=0$ in the excited state \cite{footnote}. Initially, all the atoms populate only one Zeeman sublevel, $\{F_0=1,M_0=1\}$. Two coherent control modes are applied at the empty transitions and a weak left-handed polarized probe propagates through the sample. The presence of two control modes allows us to consider the initial system as stable, otherwise with only one control field we would have to consider repopulation dynamics and optical anisotropy effects in a tripod configuration \cite{Petrosyan1,Petrosyan2}.

In the lowest order, the transfer of atoms into $\{F_0=1,M_0=-1,0\}$ sublevels is negligible and the steady state dynamics of an atomic dipole in the ensemble, as derived in \cite{SKKH}, can be expressed by the following equation for its positive frequency component $d^{(+)}(\omega)$:
\begin{eqnarray}
\lefteqn{\hspace{-0.3cm}-i\omega d^{(+)}(\omega)=-i\left[\omega_{0}+\frac{\Omega_c^2}{2(\omega-\omega_c)}\right]d^{(+)}(\omega)}%
\nonumber\\%
&&\hspace{-0.3cm} +\frac{i}{\hbar}d_0^2 \left[{\cal E}^{(+)}(\omega)+\frac{4\pi}{3}{\cal P}^{(+)}(\omega)\right]%
\!-\! i\Sigma(\omega)d^{(+)}(\omega).%
\label{2.1}
\end{eqnarray}
$\omega_0$ denotes the transition frequency, $\Omega_c$ the Rabi frequency for the control modes of frequency $\omega_c$ and $d_0=|(\mathbf{d}\cdot\mathbf{e})_{nm}|$ is the modulus of the transition matrix element between $\{n\equiv F=0,M=0\}$ and $\{m\equiv F_0=1,M_0=1\}$. The probe driving amplitude ${\cal E}^{(+)}(\omega)$ is the macroscopic transverse electric field at frequency $\omega$, considered at the point of the dipole's location and ${\cal P}^{(+)}(\omega)$ is the local mesoscopically averaged polarization. The presence of this last term in Eq. (\ref{2.1}) corresponds to the so called Lorentz-Lorenz or local field correction associated with the longitudinal (quasi-static) interaction of the dipole with its local environment. Since in the self-consistent model the proximate dipoles are indistinguishable in the excitation process, the averaged polarization can be written as
\begin{equation}
{\cal P}^{(+)}(\omega)=n_0 d^{(+)}(\omega)=\chi(\omega){\cal E}^{(+)}(\omega)%
\label{2.2}
\end{equation}
where $n_0$ is the local density and $\chi$ the macroscopic dielectric susceptibility of the sample.

The radiation losses, given by the last term in Eq. (\ref{2.1}), can be expressed as follows
\begin{eqnarray}
\Sigma(\omega)&=&\frac{\omega^2d_0^2}{\hbar^2c^2}\int\frac{d^3k}{(2\pi)^3}\,{\cal D}_{ll}^{(\bot)}(\mathbf{k},\omega)%
\nonumber\\%
&=& -\frac{i}{4}\left[\gamma_o+\gamma_e(\omega)\right]+\Delta_{\mathrm{Lamb}}%
\label{2.3}
\end{eqnarray}
It corresponds to the radiation damping created by the transverse field emitted in the scattering process and resulting into losses of the probe escaping the sample incoherently, i.e. out of its original propagation direction. The last term in the right hand side, $\Delta_{\mathrm{Lamb}}$, selects a diverging contribution to the vacuum Lamb shift, which should be renormalized and incorporated into a physical energy of the atomic transition. The integral evaluated over wave vector together with the sum over the tensor indices $l=x,y,z$ recover the spatial components of the field emitted by the dipole and overlapping with its own location. We now look for the expressions of $\gamma_o$ and $\gamma_e(\omega)$.

For an infinite medium with anisotropic dielectric susceptibility, responding only on a left-handed polarized mode, the tensor of the dielectric permittivity is given by
\begin{equation}
 \hat{\epsilon}(\omega)=\left(\begin{array}{ccc}
 1&0&0\\
 0&1+4\pi\chi(\omega)&0\\
 0&0&1
 \end{array}\right)
\label{2.4}
 \end{equation}
where the columns/rows subsequently numerate $\sigma_{+}$, $\sigma_{-}$ and $\pi$  polarizations, as shown on Fig.~\ref{fig1}. Light can propagate inside such medium into two different modes $\omega=\omega_{o/e}(\mathbf{k})$ expressed by the following dispersion equation
\begin{eqnarray}
\lefteqn{\left(c^2\mathbf{k}^2\!-\!\omega^2\right)}
\nonumber\\%
&&\times\left\{c^2\mathbf{k}^2\!-\!\omega^2[1+4\pi\chi(\omega)]\!+\!c^2(\mathbf{k}^2-k_z^2)2\pi\chi(\omega)\right\}=0%
\nonumber\\%
\label{2.5}%
\end{eqnarray}
One root is the ordinary mode $\omega=\omega_{o}(\mathbf{k})=c |\mathbf{k}|$, same as in vacuum, but another one is extraordinary and has an anisotropic dependence $\omega=\omega_{e}(\mathbf{k})$.

The transverse electric field Green's function ${\cal D}_{lm}^{(\bot)}(\mathbf{k},\omega)$ responsible for light propagation in the anisotropic medium has quite cumbersome analytical structure. But its trace (sum over $l=m=x,y,z$), which is only contributing in Eq. (\ref{2.3}), can be written in a relatively simple analytical form
\begin{eqnarray}
\lefteqn{{\cal D}_{ll}^{(\bot)}(\mathbf{k},\omega)=-\frac{4\pi\hbar}{k^2}\left\{2+\frac{\omega^2}{c^2k^2-\omega^2-i0}+\right.}%
\nonumber\\%
&&\hspace{-0.5cm}\left.+\frac{\omega^2[1+4\pi\chi(\omega)]}{c^2k^2-\omega^2[1+4\pi\chi(\omega)]+c^2(k^2-k_z^2)2\pi\chi(\omega)}\right\}
\label{2.6}
\end{eqnarray}
Evaluation of the integral leads to $\gamma_o=\gamma$, where $\gamma$ is the natural decay rate and
\begin{equation}
\gamma_e(\omega)=4\frac{d_0^2\omega^3}{\hbar c^3}\sqrt{\frac{1+4\pi\chi(\omega)}{2\pi\chi(\omega)}}\arcsin\sqrt{\frac{2\pi\chi(\omega)}{1+2\pi\chi(\omega)}}.%
\label{2.7}%
\end{equation}
We associate $\gamma_o$ with radiation emission into the ordinary mode (first denominator's pole in Eq.(\ref{2.6})) and $\gamma_e$ with emission into the extraordinary mode (second denominator's pole in Eq.(\ref{2.6})). The last value contains not only density correction of the decay rate, but also cooperative correction to the radiation Lamb shift. The key feature of this result is the existence of two different modes in anisotropic medium, namely ordinary and extraordinary, with decay rates different for the emission into each of them.

The above formulas finally enable to obtain a closed algebraic equation for the local dielectric susceptibility $\chi(\omega)$. This equation can be solved numerically and further applied for the description of the signal pulse transport through the atomic sample. In particular, for a slab sample of length $L$ filled by the medium with dielectric constant $\epsilon(\omega)=1+4\pi\chi(\omega)$ the transmission amplitude at frequency $\omega$ is given by
\begin{equation}
{\cal T}_\omega=\frac{2\sqrt{\epsilon(\omega)}}{2\sqrt{\epsilon(\omega)}\cos\psi(\omega)-i(1+\epsilon(\omega))\sin\psi(\omega)}%
\label{2.8}%
\end{equation}
where $\psi(\omega)=L\sqrt{\epsilon(\omega)}\omega/c$. The transmission coefficient $|{\cal T}_\omega|^2$ can be calculated and compared with the counterparting result of the microscopic calculations presented later in the paper.

\begin{figure}[tp]
{$\scalebox{0.7}{\includegraphics*{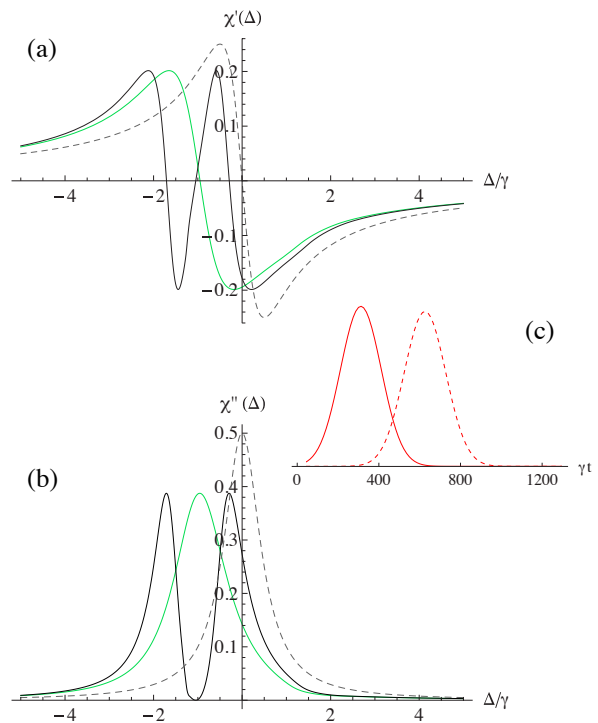}}$ }
\caption{(Color online) Spectral dependency of the dielectric susceptibility for an atomic density $n_0\lambdabar^3=1$ and electromagnetically-induced transparency features. Plot (a) provides the real part $\chi'$(dispersion) while plot (b) shows the imaginary one $\chi''$ (absorption) as function of the probe detuning $\Delta=\omega-\omega_0$. The grey dashed lines correspond to the original atomic Lorentzian profile and the green lines show how the cooperative and local field effects modify the sample susceptibility, with a resonance shift to the red.  In the presence of the control modes with Rabi frequency $\Omega_c=\gamma$, the susceptibility is strongly modified as given by the thick black lines: a transparency window and a negative dispersion appear. In this configuration, plot (c) shows the pulse propagation through an atomic array consisting of only one hundred atoms. The solid line corresponds to the initial pulse while the dashed line shows the pulse delay which achives approximately $0.5$ of the initial pulse duration and the retrival efficiency which can achive more than $50\%$ for the choosen parametes.}
\label{fig2}%
\end{figure}

As a first result, Fig.~\ref{fig2} provides the spectral dependencies of the dielectric susceptibility calculated in the self-consistent approximation for an atomic density $n_0\lambdabar^3=1$. The plots show how the original single-atom resonance profile is modified by the interatomic interactions via static longitudinal and radiation transverse fields. The resonance point is first shifted to the red wing due to the local field correction and the spectral profile also differs from its original Lorentzian shape. With control fields with $\Omega_c=\gamma$ tuned at resonance, the susceptibility exhibits typical signatures of the electromagnetically induced transparency, i.e. transparency window and reduced group velocity. In the considered case the overall resonance coupling strength, expressed by the transmission amplitude (\ref{2.8}), can attain the level required for effective pulse delay under EIT protocol, for an atomic array consisting of only one hundred atoms. This is shown in Fig.~\ref{fig2}c where the transport of a Gaussian pulse through such a short atomic array practically without losses is displayed.

We now turn to the microscopic approach.

\section{Microscopic approach}
In ab-initio microscopic quantum theory the scattering process is described by the standard $T$-matrix formalism \cite{GoldbergerWatson}, which has been adjusted for the calculation of light scattering in an ensemble of atomic dipoles in \cite{SKKH,Sheremet}. Let us consider a macroscopic target consisting of atoms randomly but homogeneously distributed in a cubic box scaled by a length $L$. Then the scattering of a plane wave mode of frequency $\omega$ is described by the following total cross section
\begin{eqnarray}
Q_0(\omega)&=&\frac{{\cal V}^2}{\hbar^2c^4}\frac{\omega'^2}{(2\pi)^2}\int\sum_{g',\mathbf{e}'}T_{g'\mathbf{e}'\mathbf{k}',g\,\mathbf{e\,k}}(E_i+i0)d\Omega'%
\nonumber\\%
&=&-\frac{2{\cal V}}{\hbar\,c}\;\mathrm{Im}\, T_{g\mathbf{e}\mathbf{k},g\,\mathbf{e\,k}}(E_i+i0)%
\label{3.1}%
\end{eqnarray}
where $T_{g'\mathbf{e}'\mathbf{k}',g\,\mathbf{e\,k}}(E_i+i0)$ are the $T$-matrix components for transition from the initial to any final state. The initial state with energy $E_i$ is specified by quantum numbers for collective atomic state $g$ (assuming all the atoms in the spin oriented state), by the mode wave vector $\mathbf{k}\parallel z$ and by its polarization $\mathbf{e}\rightarrow\sigma_{-}$, i.e. left-handed polarized. In the final state superscribed by prime sign the atomic state $g'$ can be any of the accessible ground states of the collective atomic subsystem.  The sum expands over all the allowed output scattering channels and the frequency of outgoing photon $\omega'=\omega$ for either Rayleigh or elastic Raman-type transitions shown in figure \ref{fig1}. The quantization volume ${\cal V}$ is internal and finally a vanishing parameter of the theory. It should not be confused with the target volume $V=L^3$. The second line of Eq. (\ref{3.1}), known as the optical theorem, links the total cross section with the elastic scattering amplitude in forward direction. This makes possible the calculation of the total cross section even for extremely high number of the output scattering channels $g'$. The spectral dependence $Q_0(\omega)$ describes the microscopic and configuration dependent spectral behavior of the scattering process.

For the system of atomic dipoles the $T$-matrix can be expressed by the resolvent operator (many particle Green's function) as follows \textbf{:}
\begin{eqnarray}
\lefteqn{T_{g'\mathbf{e}'\mathbf{k}',g\,\mathbf{e\,k}}(E)=\frac{2\pi\hbar\sqrt{\omega'\omega}}{{\cal V}}%
\sum_{b,a=1}^{N}}%
\nonumber\\%
&&\hspace{1 cm}(\mathbf{d}\mathbf{e}')_{nm'_b}^{*}(\mathbf{d}\mathbf{e})_{nm_a}%
\mathrm{e}^{-i\mathbf{k}'\mathbf{r}_b+i\mathbf{k}\mathbf{r}_a}%
\nonumber\\%
&&\langle\ldots m'_{b-1},n,m'_{b+1}\ldots |\tilde{\hat{R}}(E)%
|\ldots m_{a-1},n,m_{a+1}\ldots \rangle%
\nonumber\\%
&&\label{3.2}%
\end{eqnarray}
where the resolvent $\tilde{\hat{R}}(E)=P(H-E)^{-1}P$ is projected by the operator $P$ onto field vacuum state and onto a set of atomic states with a single optical excitation between the states $m$ and $n$. $H$ is the total system Hamiltonian in the dipole gauge \cite{ChTnDRGr}. Indices $a$ and $b$ numerate the atoms and indicate their locations and occupation states. As in the previous section we assume that the excited atomic level has only one state $n$, but the atoms can be repopulated among all the ground state Zeeman sublevels $m$ in the interaction process. The resolvent can be calculated numerically as shown in Ref. \cite{SKKH,Sheremet} and the presence of the control field can be incorporated into its self-energy part via similar term as in the right-hand side of Eq. (\ref{2.1}).

The basic theory and supplementary details of the microscopic calculation scheme can be found in \cite{SKKH,Sheremet}. Generally for a large atomic ensemble consisting of $N$  atoms with degenerate ground state the exact resolvent operator is defined in an Hilbert subspace of large dimension, which is exponentially rising up with $N$ as $d_e N d_g^{N-1}$. Here $d_g$ and $d_e$ are the degeneracies of the ground and excited states respectively. This makes the dimension of the Hilbert subspace for macroscopic $N\gg 1$ tremendously large for degenerate system when $d_g\neq 1$ as in our case. Nevertheless by evaluating the total cross-section via the optical theorem \ref{3.1} we need to know only the diagonal elements of the $T$-matrix associated with elastic forward scattering. As a consequence, for the resolvent operator it is necessary to only access a limited set of its matrix elements such as defined in the subspace where $N-1$ atoms are in the same ground state and a single excitation is shared among any of $N$ atoms of the ensemble. In the Feynman diagram expansion the resolvent is expressed by $N$-particle Green's function obeying the collective Dyson equation \cite{Sheremet}. The irreducible diagrams form so-called self-energy part $\Sigma(E)$ and finally the inverse resolvent operator can be expressed as follows
\begin{equation}
\tilde{\hat{R}}^{-1}(E)=E-\hbar\omega_0-\Sigma(E).%
\label{3.3}
\end{equation}
While calculating the self-energy matrix we apply the standard pole or "on-shell" approximation, similar to the Wigner-Weiskopf approach for single atom spontaneous emission, and substitute in it $E=\hbar\omega_0$, i.e. the undressed energy of the excited state with the assumption for the ground state energy $E_g=0$: $\Sigma(E)\to \Sigma(\hbar\omega_0)$.

The cooperative contribution to the self-energy part basically consists of the vacuum photon propagators linking the transition currents of any atomic pairs in the ensemble and we readdress the reader to \cite{Sheremet}, where the collective Dyson equation as well as the self-energy operator are discussed. The formalism is applicable for the conditions fulfilling the dipole gauge approximation i. e. for the densities up to $n_0\lambdabar^3\sim O(1)$ and not higher. Then, as shown in \cite{Sheremet}, the matrix element of the self-energy operator responsible for the coupling of atoms $a$ and $b$ is given by
\begin{equation}
\Sigma^{(ab)}_{m'n;nm}(\hbar\omega_0)=
\frac{1}{\hbar}\,d^{\mu}_{nm}d^{\nu}_{m'n}\,D^{(E)}_{\mu\nu}(\mathbf{R}_{ab},\omega_0).%
\label{3.4}%
\end{equation}
where the causal-type Green's function of the microscopic displacement field $D^{(E)}_{\mu\nu}(\mathbf{R},\omega)$ is taken at the distance of interatomic separation $\mathbf{R}=\mathbf{R}_{ab}$.

Such a relatively simple structure of the self-energy part allows us to select a sequence of diagram, which gives main contribution into the diagram expansion of the resolvent operator, and then keep only these diagrams in the entire calculation scheme. The basic principle can be explained in the lowest order of the diagram expansion. As example in the second order of the $\tilde{\hat{R}}(E)$ expansion in the series of the self-energy part we select the following contribution
\begin{equation}
\scalebox{0.6}{\includegraphics*{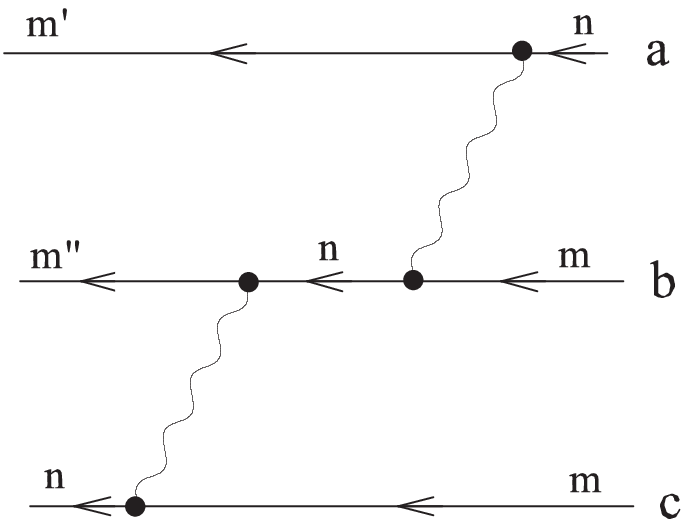}}
\label{3.5}
\end{equation}
where $a,b,c=1\div N$ are any of atoms in the ensemble. Having in mind application of the optical theorem and considering this contribution to the elastic scattering channel all the ground state quantum numbers should coincide, i. e. $m''=m'=m$. For large $N\to\infty$ because of statistical enhancement such type of diagram contributions, where different wavy lines link only different atomic lines, will dominate in the main sequence of perturbation theory expansion for the resolvent operator. For contributions of this type we have to keep only the swapping components ($n,m\to m,n$) of the self-energy operator for each interacting pair.

Let us now specify another diagram also contributing in the second order of the perturbation theory
\begin{equation}
\scalebox{0.6}{\includegraphics*{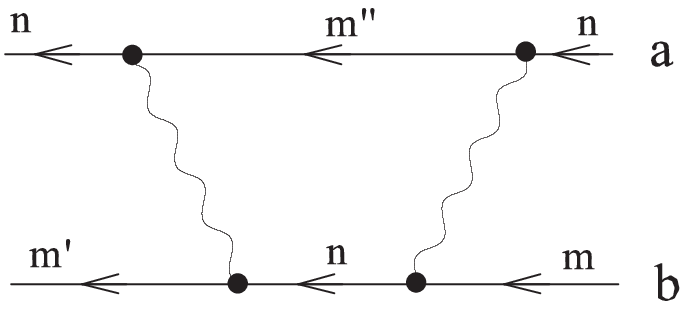}}
\label{3.6}
\end{equation}
This diagram shows us an example of recurrent coupling and virtual repopulation among two neighboring atoms. For this diagram to contribute to elastic Rayleigh channel, we have $m'=m$ but $m''\neq m$. Diagram (\ref{3.6}), as well as similar graphs appearing in higher orders, can be added into any inner part of the main diagram sequence starting by (\ref{3.5}) and expanded up to higher orders. As a consequence the elastic channel becomes coupled with other Raman-type scattering channels, which are elastic in terms of conserving energy but having different magnetic numbers, and for the sake of simplicity in definitions we will further address them as inelastic.

From a more physical point of view, the recurrent virtual processes are responsible for cooperative corrections to a single atomic excitation from both the static longitudinal and radiation transverse fields modified by the presence of proximal neighbors. For intermediate densities, $n_0\lambdabar^3\sim 1$, it would be sufficient to keep such recurrent coupling only for the atoms separated by a distance of few $\lambdabar$. This can essentially reduce the number of inelastic scattering channels coupled with the elastic one and as a consequence the subspace dimension for the resolvent operator as well as the number of equations to be solved, which would scale now as $d_e N d_g^{\mathrm{n}-1}$. Here "$\mathrm{n}-1$" corresponds to the effective number of the proximate neighbors, which mainly contributes in the recurrent diagrams responsible for "dressing" of any randomly selected atomic excitation in the ensemble. In the calculations this number is varied such that we can verify that the entire calculation scheme with increasing "$\mathrm{n}$" becomes rapidly self-converging and in the limit of $\mathrm{n}\gg n_0\lambdabar^3$ it approaches the exact result.

\section{Results}

We now investigate how the effect of disorder, associated with the dense random distribution of atomic scatterers, can affect the result of the self-consistent description and make absorption profile potentially sensitive to a specific atomic configuration.

The transmission coefficient $|{\cal T}_\omega|^2$, defined by  Eq.(\ref{2.8}), and the spectral profile of the scattering cross section $Q_0(\omega)$, given by Eq. (\ref{3.2}), are different quantities. Nevertheless in classical theory of diffraction, in accordance with the Babinet's principle, the light scattered by highly absorbing macroscopic sample has its scattering cross section equal to $2{\cal A}$, where ${\cal A}$ is the geometrical cross-area of the object. In the vicinity of the absorption resonance we can thus expect the approximate relation $2{\cal A}[1-|{\cal T}_\omega|^2]\sim Q_0(\omega)$, which can be used to test the validity of the self-consistent macroscopic description via round of microscopic calculations. In the spectral domain where the sample becomes partially transparent both calculation schemes approach the same limit of weak light scattering on a collection of independent point-like atomic dipoles. In this limit we have $Q_0(\omega)=N\sigma(\omega)$, where $N=n_0V$ is the number of scatterers and $\sigma(\omega)$ is the cross section of light scattering on a single atom. In the case of weak interaction the latter quantity also contributes to the susceptibility $4\pi k\,\mathrm{Im}\chi(\omega)=n_0\sigma(\omega)$ such that we get asymptotic relation between both calculation approaches.

\begin{figure}[tp]
{$\scalebox{0.9}{\includegraphics*{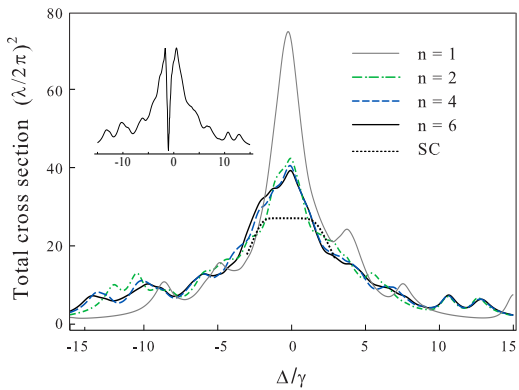}}$}
\caption{(Color online) Total cross section for a single photon scattering on an ensemble consisting of fifty atoms randomly distributed with a density $n_0\lambdabar^3\sim 1$. The microscopic calculations have been performed for a particular configuration and for different numbers "$\mathrm{n}$" of proximate neighbors involved in the recurrent coupling, as described in the text. The inset shows the microscopic verification of EIT interaction with control pulses calculated for the same parameters as used in Fig. \ref{fig2}. The self-consistent (SC, dotted line) estimate of the cross section spectrum (black dotted) is scaled in accordance with the Babinet's principle and only the part of maximal absorption is shown, see text.}
\label{fig3}%
\end{figure}

The spectral behavior of the scattering cross section $Q_0(\omega)$ is shown in Fig.~\ref{fig3} for an ensemble consisting of fifty atoms with a density $n_0\lambdabar^3\sim 1$ and for a number of proximate neighbors contributing in the recurrent coupling up to five. The plots, presented for different "$\mathrm{n}$", show the self-convergency of our calculation scheme with increasing "$\mathrm{n}$". The calculations were done for a particular configuration and the spectrum reproduces a randomly created quasi-energy resonance structure of the resolvent operator for the chosen configuration. The absorption profile is realistically reproducible by the self-consistent calculation scheme and variations in the atomic configuration only slightly affects the spectral behavior of the cross section near the central resonance. The EIT phenomenon can be involved in the microscopic calculation scheme by adding the coupling with the control mode, contributed to the first line of Eq.(\ref{2.1}), as a self-energy part to the atomic propagators of unoccupied states. Then the transparency window, shown in inset and calculated for the same parameters as in Fig.~\ref{fig2}, is also little sensitive to a particular atomic configuration.

As a reference dependence we have shown here a fragment of the transmission spectrum calculated in the self-consistent approximation near the resonance point and applied the Babinet's principle for the total cross section estimate at resonance. Both calculation schemes are in agreement in their general behavior and the microscopic result has a clear signature of the local field correction being slightly asymmetric to the red wing. The fact that the microscopic cross section is a bit larger at the point of maximum indicates that the macroscopic description is not so straightforwardly applicable to system of mesoscopic size. The discrepancy between micro and macroscopic estimates of the cross section can be explained as the sample size $L={}^{3}\!\!\!{\surd}50\, \lambdabar$ hits the region $\lambdabar < L<\lambda$, so it is not large enough to follow the Fraunhofer diffraction and the Babinet's benchmark. Unfortunately it is  demanding so far to extend the applied algorithm here up to macroscopic object consisting of a larger number of atoms at high density and overcoming the conditions $L>\lambda$ as it was done in \cite{SKKH} for $V$-type two level atoms.

\begin{figure}[tp]
{$\scalebox{1.0}{\includegraphics*{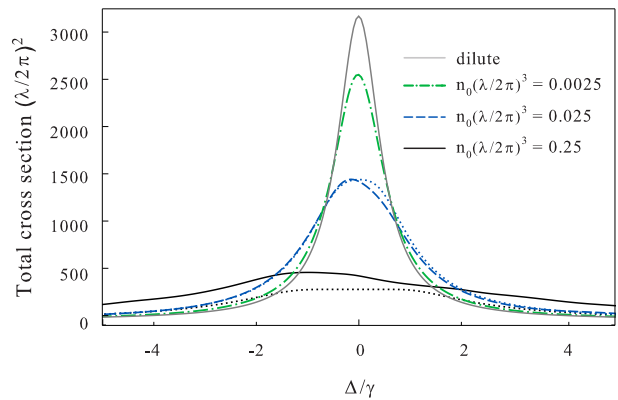}}$}
\caption{(Color online) These graphs show how the self-consistent model fits the results of ab-initio microscopic calculations in the case of macroscopic limit. All the graphs reproduce the calculations of the total cross section for a single photon scattering on an ensemble consisting of five hundreds atoms randomly distributed with the different densities. The ab-initio calculations are performed for the the densities $n_0\lambdabar^3 = 0.25$ (black), $n_0\lambdabar^3 = 0.025$ (blue dashed) and $n_0\lambdabar^3 = 0.0025$ (green dashed-dotted) and for a represenative number of proximal neighbors $\mathrm{n-1}=3$. The self-consistent estimate of the cross section spectra, shown in dotted curves with same colors for $n_0\lambdabar^3 = 0.25, 0.025$, are scaled in accordance with the Babinet's principle. The upper bounding curve (gray) indicates the cross section for the upper dilute limit when the total cross-section is directly given by the sum of individual cross-sections such that $Q_0(\omega)=N\,\sigma(\omega)$. }
\label{fig4}%
\end{figure}

By considering the evolution from a highly dense to a more dilute configurations we can justify the validity of the ab-initio microscopic and self-consistent macroscopic approaches and compromise them in accordance with the principles described in the beginning of this section. In Fig.~\ref{fig4} we present our calculations performed for an atomic ensemble consisting of $N=500$ atoms with a fixed number $\mathrm{n}=4$, i.e. keeping only three proximal neighbors. The upper bounding curve (gray) indicates the cross section in the very dilute limit when the total cross-section is directly given by the sum of individual cross-sections such that $Q_0(\omega)=N\,\sigma(\omega)$. The calculations, performed for different densities $n_0\lambdabar^3 = 0.0025, 0.025, 0.25$, show how the total scattering cross-section reduces from the dilute limit to the dependent scattering regime. At the intermediate densities for a sample of macroscopic size ($L>\lambda$) a good agreement is obtained between both calculation approaches. The Babinet's benchmark, shown with dotted curves, faithfully fits the results of microscopic calculations for the densities $n_0\lambdabar^3 = 0.025, 0.25$.

\section{Conclusion}

In conclusion, we have shown that in a dense and disordered atomic ensemble the effective slow-light transport can be obtained in atomic array with a relatively small number of atoms compared with dilute systems. However, the effect of disorder, cooperative scattering and static interaction could be expected to randomize the transmission spectrum. Interestingly, our results obtained with the ab-initio microscopic approach reveal actually that the total scattering cross-section has only slight signature of the configuration dependence. The scattering spectra demonstrate indeed a tendency to self-averaging and smooth behavior near the main resonance peak, which is in turn reliably reproducible by the self-consistent model. As a consequence this allows effective coherent control of a signal pulse under the regime of EIT with a fixed set of external parameters, such as Rabi frequency and control mode detuning, irrespective of any random realizations of the atomic configuration. Extension of this study to one-dimensional array, such as obtained with atoms trapped along an optical nanofiber where disorder can manifest due to limited filling factor, is in progress.

\acknowledgments

This work was supported by the CNRS-RFBR collaboration (CNRS 6054 and RFBR 12-02-91056), RFBR grant 15-02-01060 and the Emergence Program from Ville de Paris. A.Sh. acknowledges fellowship from the Russian President Program and is supported by the Program 5-100-2020 of SPbSPU. D.V.K. would like to acknowledge support from the External Fellowship Program of the Russian Quantum Center and L.V.G. from the Foundation "Dynasty" and the Alferov's Foundation. J. L. is a member of the Institut Universitaire de France.


\begin{thebibliography}{10}
\bibitem{footnote} The considered configuration of atomic levels is not exactly what is usually called a $\Lambda$-type scheme based only on three levels. But we emphasize that there is no strict example of three-level atom in reality. The energy level scheme that we study here seems the closest to the $\Lambda$-type that one can encounter in real physical systems.
\end{thebibliography}
\end{document}